# Employee-Driven Innovation to Fuel Internal Software Startups: Preliminary Findings

Anastasiia Tkalich[1(✉)][0000-0001-7391-4194], Nils Brede Moe[2][0000-0003-2669-0778]

Tor Sporsem [3][0000-0002-5230-7480]

[1,2,3] SINTEF Digital, 7034 Trondheim, Norway
[1]anastasiia.tkalich@sintef.no

**Abstract.** To keep up with the pace of innovation, established companies are increasingly relying on internal software startups. However, succeeding with such startups is a challenging task because internal startups need to find a balance between the interests of the company and the interest of the innovator. One approach that is argued to strengthen innovation in existing companies is employee-driven innovation (EDI). This study explores this argument by examining two internal software startups in companies aligned with the principles of EDI and with a strong focus on innovation. The preliminary findings indicate that startups with EDI are characterized by commitment towards innovation, cooperative orientation, and autonomy. The findings suggest that internal software startups may be strengthened when the parent companies practice EDI.

**Keywords:** Internal Software Startup, Internal Startup, Software Product Innovation, Employee-Driven Innovation, Software-Intensive Business.

## 1    Introduction

Building a successful software startup is difficult, and very few can copy the success stories of companies like Uber, Airbnb, Spotify, and Zoom. Many software startups fail before they reach their commercial potential [1]. In order to aggressively grow and scale, they need to balance high-speed innovation in extreme uncertainty with issues such as building entrepreneurial teams, acquiring a customer base, and operate in a sustainable way [2]. Innovation is also the key for established software-intensive companies. To stay competitive, they are increasingly adopting startup approaches within their own environment, thus challenging the traditional view of startups as being independent. Such internal startups [3] offer big companies flexibility and growth that can potentially boost their software product innovation. 82% of respondents in 170 large companies reported using some elements of the startup approach to accelerate their innovation [4].

However, driving internal software startups is challenging. Comparing to regular startups, the leaders of internal startups do not always have a personal stake in the final results, which makes it hard to maintain their motivation [3]. The startups may be disconnected from other crucial parts of the organizations, such as software development units, something that may hinder coordination and thus render the development process unstable [5]. Finally, being a part of a larger corporation typically reduces the innovator´s autonomy because they need to take into account not only external customers but also the corporate management [6]. In such a situation, the innovators may lose the freedom to pivot and experiment, which is considered crucial for all startups [3].

One approach that is argued to strengthen innovation within existing companies is known as employee-driven innovation (EDI) [7]. EDI assumes that all employees have skills and experience that can increase the organization's overall capacity to innovate under favorable conditions, which is also aligned with one of the principles of Lean Startup "Entrepreneurs are everywhere" [8]. Today´s practices in software product innovation imply that employees take an active role in driving new products [9]. One can thus expect that companies with the increased role of employees in innovation will also be more successful in internal software startups. Despite the seeming potential of EDI to strengthen internal software startups, research on this topic is only starting to emerge in software engineering. We, therefore, ask the following research question: *How do companies with employee-driven innovation (EDI) drive their internal software startups?* To answer, we are reporting preliminary findings from data collected in two Norwegian companies with EDI and with a strong focus on innovation.

## 2  Related work

Traditionally, startups are understood as temporary organizations focused on innovative products with little or no operative history, aiming to grow by aggressively scaling their businesses [10]. With the increasing availability of software usage and the influence of Lean Startups, software startups have become more and more widespread [11]. They rapidly find scalable business models, build new products, and pivot if necessary, which creates pressure for established software-intensive companies.

As a response, the established companies are increasingly adopting so-called internal software startups. Just as regular (external) startups, internal software startups are risk-taking and proactive initiatives that develop software under highly uncertain conditions by constantly searching for repeatable and scalable business models [11, 12]. The main difference is, however, that internal software startups are nested within the existing parent companies [6]. Unlike external startups that rely on limited resources, internal startups benefit from several existing resources (e.g., salary) [13]. Further, if the external startups are mainly driven by the needs of their external customers, the internal startups also must consider their corporate management [13]. However, various types of internal startups may differ in how much they depend on their parent company. For example, internal ventures are typically entirely supported by the resources from the parent company. In contrast, spin-offs are based on the technology from the parent company but do not entirely rely on it for the resources and eventually become independent new companies [14].

To fuel the development of new products and businesses through internal software startups, companies often rely on specific innovation strategies [15]. For example,

3
Google relies on practices like design sprints [9], where developers are given one day to showcase a proof of concept they believe should be part of the product, whereas Atlassian applies the method "20% Time" to allow more ambitious innovation projects to be undertaken [15]. One common trait in the practices described above is that the employees' own initiative and entrepreneurial skills play a central role for the internal startups, which is characteristic of *employee-driven innovation* (EDI) [7]. EDI is seen as an informal innovation process [16], meaning that the employees are not formally assigned to innovation tasks. Therefore the EDI approach is different from traditional research and development (R&D) when novel products are developed by a dedicated formalized unit [7]. In contrast, the EDI approach allows the innovation to be driven by employees from any unit, and it is up to the employee whether to engage in the process or not.

Organizations that systematically engage in EDI seem to share a set of cultural characteristics. Based on a study of 20 companies, Aasen et al. [16] have identified nine shared features: commitment, cooperative orientation, pride, trust, tolerance, feeling of security, development orientation, openness, and autonomy. In this study, we specifically focus on the three features that were central in our data: *commitment*, *cooperative orientation,* and *autonomy* (see Table 1 for the definitions).

**Table 1.** Cultural characteristics of employee-driven innovation [16]

| Cultural characteristics | Description |
| --- | --- |
| Commitment | High commitment towards innovation among employees |
| Cooperative orientation | Basic assumption that there is agreement to cooperate between management and employees |
| Autonomy | Employees have a high degree of influence in relation to the execution of various tasks |

Even though it has been described how EDI is practiced in other contexts [16] and the term has been adopted by the field of Information Systems [17], it is not sufficiently studied in software engineering. Our study seeks to address this knowledge gap by examining the companies that are both aligned with the principles of EDI and focus on innovating through internal software startups.

## 3  Methods

To answer the RQ, we collected data from two cases of internal software startups in two Norwegian companies (Table 2). We selected companies with internal software startups and where employees were first involved in the innovation process by taking an informal innovator role on their own initiative (EDI). Both companies are technology-intensive with a strong focus on innovation. The first company is Iterate that was listed among the 100 best workplaces for innovators in 2020 [18]. The second company is MarComp (real name suppressed for anonymity), which has since 2018 been actively training personnel and management in software product innovation. In both companies, we conducted semi-structured interviews and collected documents and meeting notes between September 2020 and April 2021. Different sets of questions were used to interview product managers comparing to interviewing other stakeholders (e.g., CEO, innovation facilitators). Examples of the questions for product managers (employees in charge of the startups) are "*Please, tell us about your innovation project and how it arose,*" "*Who has been the most important collaboration partner?*", "*What has been the biggest challenge, and how did you solve it?*". The interview guide for the other stakeholders consisted of questions like "*How does your company captures ideas for new internal startups?*", "*How do you decide which ideas are good enough?*". In terms of the documents, we collected slides, emails, and websites that reflected the nature of the startups, their history, and the organizational context they operated in.

**Table 2.** Data sources

| Parent firm | Industry | Startup | Data sources | # | Details |
|---|---|---|---|---|---|
| Iterate (Norway) | Venture-builder/IT | CalcTool | Interviews | 4 | Product manager, CEO, Designer, Customer |
| | | | Meeting notes | 2 | Informal meetings with CEO |
| | | | Documents | | Website, product strategies (slides), emails |
| MarComp | Maritime | ShipDash | Interviews | 3 | Product manager (two times), innovation facilitator |
| | | | Meeting notes | 5 | Community of practice |
| | | | Documents | | Innovation program's slides, emails |

For the current analysis, we first created an overview of the cases based on the contextual characteristics that could be compared across the cases. Then we identified the data instances that had to with how the parent companies drove the startups. Finally, we grouped the innovation strategies thematically in an attempt to categorize different types of strategies. The data were analyzed by the first author, who continuously consulted the other authors to validate the emerging results. The earlier version of the paper was also shared with the key interviewees to collect their feedback, which was later incorporated in the final version.

## 4      Results

We will describe the internal software startups and how the parent companies drove them (innovation strategies). Key contextual characteristics of the case startups are presented in Table 3.

**Table 3.** Overview of the internal software startups

| Contextual characteristics | CalcTool | ShipDash |
|---|---|---|
| Size of the parent company | S | L |
| Beginning of the startup job | Q3,2020 | Q4 2019 |
| Type of internal startup | Spin-off | Internal venture |
| Dedicated employee (product manager) | ✓ | ✓ |
| The startup job is financed by the parent company | Partially | ✓ |
| The product manager validates strategic decisions with the parent company | | ✓ |
| Size of the startup team | 5 | 6 |
| Innovation coaching | ✓ | ✓ |

*Note.* ✓ = the characteristic is identified in the startup

**CalcTool.** The aim of the startup is to create a new software-enabled online tool for construction engineers. Iterate (the parent company) functions as a venture builder investing in software startups and additionally provides IT consultancy services. Apart from the product manager, the CalcTool team consists of two software developers and two designers from the same company who volunteered to join the initiative. Interested in the idea, the team members agreed to use their own hour-budget sponsored by Iterate ("Iterate time") to work on the startup. The product manager expressed: "*I tried to recruit others and convince them to donate their Iterate time to me.*" The team closely collaborates with potential users (construction engineers) who provide input on the desired functionality. The product manager emphasized: "*It is very beneficial to be a part of the construction engineer network.*" The team's intention is to create an independent spin-off where Iterate can potentially become an investor. The team has full autonomy about the technical and strategic decisions with no influence from Iterate. The startup's product manager is a full-time employee at the parent company and works with the startup only 10 hours per month in addition to his spare time. He commented: "*I am willing to invest my time in this because I can get a share in the company that can come out.*" Although Iterate is only partially financing the startup job through "Iterate time," this job is very encouraged by the CEO. The startup receives support from Iterate in the form of:

- coaching - whenever the product manager requests it, the top manager acts as a coach. This is a way for the managers to be involved in the startup without taking too much control. The CEO commented: "*If people feel that they have to validate things with us [...] it will go too slow. But when there is something to show, we are glad to be in a dialog*";
- networks – there is an external Slack channel on product management that is actively used by many other employees inside and outside the company and is part of the informal project management school driven by the executive manager. Earlier, the product manager worked in another startup team in the parent company, which gave him experience in the startup work. He emphasized the importance of collaborating with other employees: "*It gives me confidence in being as I am.*";
- ceremonies - examples of the ceremonies are 1) breakfast meetings (employees meet from 7.30 to 8.30 on certain days for pitching their entrepreneur ideas); 2) "While we wait" conference on Wednesdays (11-12) where most experienced employees present their current startups; 3) "Ship-it day" – A 24-hour hackathon that can (once per year) be used for delivery in one's own project;

- culture – innovation mindset is part of the organizational culture. Individual employees are encouraged to create their own software startups and collaborate with each other to spur creativity, experience, and motivation. A designer from another project expressed: "*In Iterate, we have a shared culture that helps us collaborate even if we never worked with each other before.*"

**ShipDash** is developing an algorithm-enabled tool for insight into ships' emissions based on various data sources (e.g., AIS, reported fuel data from vessels). The startup is a part of an internal innovation program, and in November 2020, moved to the operational department for scaling. Since the startup is entirely supported by the resources from the parent company, it can be categorized as internal venture [14]. Key strategic decisions (e.g., scaling, extra financing, brand name) are validated with the parent company, which can slow down the development process. The product manager confessed: "*It is so challenging sometimes, because there are so many [managers] who have an opinion on what we are building should be.*" However, the innovation facilitators are constantly working on expanding the product manager´s autonomy in the firm, thus enabling faster user testing and pivoting. For example, sub-branding was introduced to lower the threshold for the product managers to test their solutions with the existing customers without harming the main company brand. The product manager of ShipDash said: "*Sub-branding is cool; it makes me want to innovate more […]. Now I can have a label [towards customers] that I am just testing this and learning*". Since August 2020 (building and scaling), the product manager is working on the startup most of her time (85% on average), but in the past, the workload varied from 20% to 50%). This became possible because her line manager accepted her working less on the main position in favor of the startup. The parent company supports the startup by:

- coaching – provided by both in-house and external innovation coaches and internal mentors (experienced leaders, domain experts, innovation facilitators); The coaches and innovation facilitators also function as problem-solvers when the product managers are stuck. One product manager commented during a meeting: "*If we said, oh we need to have more customer contacts, then the coach said Ok, I will ask this person, and he will organize something*";
- networks – 1) customer relations managers gave access to test-users during the experimentation phase; 2) stakeholders from marketing, customer relations, finance, operations, and software were frequently invited to take part in strategic and technical decisions on both periodic (ceremonies) and non-periodic arenas (meeting requests): 3) informal networks with other product managers who shared their experience from similar startups were also crucial. The product manager emphasized: "*I relied so much on my good colleagues who have been through similar projects. […] I asked one of them so many questions that he is now calling himself my mentor*";
- ceremonies – 1) Periodic "innovation board" meetings where domain experts evaluate the startup's progress; 2) "innovation guild" – meetings of a community of practice driven by other product managers and innovation coaches to share experiences and drive the internal startups;
- processes - using a stage-gate innovation process based on the Lean Startup approach, the company encourages the employees to suggest ideas (Call for ideas) and then develop a minimal viable product with high potential for scalability. The innovation process also implies that the product managers have the freedom to conduct user testing with the customers as part of the product development.

## 5      Discussion and practical implications

To answer the research question *How do companies with employee-driven innovation (EDI) drive their internal software startups?* we have presented preliminary findings from two cases of startups in companies with EDI. In both cases, the employees were working with the startups only part-time, but the workload seemed to vary and generally increased along with the startup's maturity. Both parent companies were relying on a set of innovation strategies to motivate their employees to engage in internal startups, such as coaching, networks, and ceremonies. Whereas the larger company had a stronger focus on how to structure the innovation processes, the smaller company focused more on shared innovation culture. We will now discuss our findings against the cultural features of EDI and earlier research on internal startups to understand the differences between internal startups with and without EDI. Due to the limited format of the workshop paper, we focus only on the three cultural features of EDI that were particularly central in our findings: *commitment*, *cooperative orientation,* and *autonomy* [16].

Our case startups were characterized by a strong commitment to the innovation process and outcome. The product manager of CalcTool was willing to work on the startup in his spare time. The startup team contained other employees that were committed to due to the shared innovation mindset in the company. In ShipDash the line managers showed their commitment to the innovation process by reducing the product manager workload, which freed them for the startup work. As a result, the product manager was willing to take up a new role in the company, as the startup was gradually occupying most of her work time. Earlier research on internal startups demonstrated that maintaining the motivation of employees to innovate is limiting for internal startups [3]. Our findings indicate that this problem can be solved by EDI that creates conditions where both employees and managers are motivated to contribute to internal software startups.

Our results suggest that companies with EDI are actively promoting cooperation not only between employees and management (as described earlier [16]) but also between employees of different departments and externally with customers. Iterate was using a plethora of ceremonies to promote networking among the employees where they could pitch their ideas, receive feedback from more experienced innovators and recruit others to the startup team. In ShipDash the parent company was also promoting the product managers´ networks with key customers, domain experts (marketing, finance, software), and more experienced product managers. Earlier research indicated that access to existing networks of experts internally in the company is enabling for the internal startups [3]. External networks with customers or corporate partners are also important for startups in general because they tend to depend on the innovation ecosystem around [2]. Based on our findings, we can thus conclude that organizations with EDI can have a positive effect on internal startups by promoting networks both internally and externally.

Finally, our findings indicate that achieving autonomy is central for internal software startups in EDI-oriented companies but that it can also be challenged. Specifically, CalcTool enjoyed high autonomy and did not have to validate the strategic decisions with the management, whereas in ShipDash the product manager´s autonomy was lower. However, MarComp was systematically working to increase the autonomy of the innovators (e.g., freedom to experiment, introduction of sub-branding). Autonomy in the decision-making process is crucial for internal startups because it speeds up development and learning through experimenting [3]. Even though EDI-companies can differ in the degree of the startups´ autonomy (due to differences in the companies´ size, financial or legal concerns), such companies still acknowledge the importance of it and actively work to increase it by different means.

## 6      Conclusions, limitations, and future work

To gain more knowledge on internal software startups in companies with employee-driven innovation (EDI), we have reported preliminary findings from two startups in different companies. We found that both startups were characterized by high commitment towards innovation, cooperative orientation, and autonomy, which we argued is positive for internal software startups in general. One of the central limitations of this preliminary study is that the data originates from Norwegian companies where the tradition of employee involvement is historically strong [16]. In future work, we intend to collect data from other countries and conduct an even more rigorous analysis of a larger number of startups to acquire additional insight into the effect of EDI.

**Acknowledgments.** This research was supported by the 10xTeams project and the Research Council of Norway through grant 309344.

## References

1.   Crowne, M.: Why software product startups fail and what to do about it. Evolution of software product development in startup companies. In: IEEE International Engineering Management Conference. pp. 338–343. IEEE (2002).
2.   Abrahamsson, P., Bosch, J., Brinkkemper, S., Mädche, A.: Software Business, Platforms, and Ecosystems: Fundamentals of Software Production Research (Dagstuhl Seminar 18182). Schloss Dagstuhl - Leibniz-Zentrum fuer Informatik GmbH, Wadern/Saarbruecken, Germany (2018).
3.   Edison, H., Smørsgård, N.M., Wang, X., Abrahamsson, P.: Lean Internal Startups for Software Product Innovation in Large Companies: Enablers and Inhibitors. J. Syst. Softw. 135, 69–87 (2018). https://doi.org/10.1016/j.jss.2017.09.034.
4.   Kirsner, S.: The Barriers Big Companies Face When They Try to Act Like Lean Startups, https://hbr.org/2016/08/the-barriers-big-companies-face-when-they-try-to-act-like-lean-startups, (2016).

5. Sporsem, T., Tkalich, A., Moe, N.B., Mikalsen, M.: Understanding Barriers to Internal Startups in Large Organizations: Evidence from a Globally Distributed Company. In: Proceedings of 16th ACM/IEEE International Conference on Global Software Engineering (ICGSE). , Virtual conference (2021).
6. Edison, H., Wang, X., Jabangwe, R., Abrahamsson, P.: Innovation Initiatives in Large Software Companies: A Systematic Mapping Study. Inf. Softw. Technol. 95, 1–14 (2018). https://doi.org/10.1016/j.infsof.2017.12.007.
7. Høyrup, S.: Employee-driven innovation and workplace learning: basic concepts, approaches and themes. Transf. Eur. Rev. Labour Res. 16, 143–154 (2010). https://doi.org/10.1177/1024258910364102.
8. Ries, E.: The lean startup: How today's entrepreneurs use continuous innovation to create radically successful businesses. Currency (2011).
9. The Design Sprint — GV, http://www.gv.com/sprint, last accessed 2021/01/08.
10. Giardino, C., Wang, X., Abrahamsson, P.: Why Early-Stage Software Startups Fail: A Behavioral Framework. In: Lassenius, C. and Smolander, K. (eds.) Software Business. Towards Continuous Value Delivery. pp. 27–41. Springer International Publishing, Cham (2014). https://doi.org/10.1007/978-3-319-08738-2_3.
11. Paternoster, N., Giardino, C., Unterkalmsteiner, M., Gorschek, T., Abrahamsson, P.: Software development in startup companies: A systematic mapping study. Inf. Softw. Technol. 56, 1200–1218 (2014). https://doi.org/10.1016/j.infsof.2014.04.014.
12. Melegati, J., Guerra, E., Wang, X.: Understanding Hypotheses Engineering in Software Startups through a Gray Literature Review. Inf. Softw. Technol. 133, 106465 (2021). https://doi.org/10.1016/j.infsof.2020.106465.
13. Edison, H., Wang, X., Abrahamsson, P.: Lean startup: why large software companies should care. In: Scientific Workshop Proceedings of the XP2015. pp. 1–7. Association for Computing Machinery, New York, NY, USA (2015). https://doi.org/10.1145/2764979.2764981.
14. Edison, H., Wang, X., Jabangwe, R., Abrahamsson, P.: Innovation Initiatives in Large Software Companies: A Systematic Mapping Study. Inf. Softw. Technol. 95, 1–14 (2018). https://doi.org/10.1016/j.infsof.2017.12.007.
15. Moe, N.B., Barney, S., Aurum, A., Khurum, M., Wohlin, C., Barney, H.T., Gorschek, T., Winata, M.: Fostering and Sustaining Innovation in a Fast Growing Agile Company. In: Dieste, O., Jedlitschka, A., and Juristo, N. (eds.) Product-Focused Software Process Improvement. pp. 160–174. Springer, Berlin, Heidelberg (2012). https://doi.org/10.1007/978-3-642-31063-8_13.
16. Aasen, T.M., Amundsen, O., Gressgård, L.J., Hansen, K.: Employee-driven innovation in practice - Promoting learning and collaborative innovation by tapping into diverse knowledge resources. LifeLong Learn. Eur. 4, (2012).
17. Opland, L., Jaccheri, L., Pappas, I., Engesmo, J.: Utilising the innovation potential - a systematic literature review on employee-driven digital innovation. In: 29th European Conference on Information Systems (2021).
18. Fast Company: Best Workplaces for Innovators 2020, https://www.fastcompany.com/90527870/best-workplaces-for-innovators-2020, last accessed 2021/07/08.